# Project X ICD-2 and its Upgrades for Neutrino Factory or Muon Collider

Valeri Lebedev and Sergei Nagaitsev

*Fermilab, PO Box 500, Batavia, IL 60563, USA*

**Abstract.** This paper reviews the Initial Configuration Document for Fermilab's Project X and considers its possible upgrades for neutrino factory or muon collider.

## INTRODUCTION

Project X is a high intensity proton facility conceived to support a world-leading program in neutrino and flavor physics over the next two decades at Fermilab. Project X is an integral part of the Fermilab Roadmap as described in the Fermilab Steering Group Report of August 2007 [1] and of the Intensity Frontier science program described in the P5 report of May 2008 [2].

The primary elements of the research program to be supported by Project X include:

- *A neutrino beam for long baseline neutrino oscillation experiments.* A new 2 megawatt proton source with proton energies between 60 and 120 GeV would produce intense neutrino beams, directed toward a large detector located in a distant underground laboratory.
- *Kaon and muon based precision experiments running simultaneously with the neutrino program.* These could include a world leading muon-to-electron conversion experiment (mu2e) and world leading rare kaon decay experiments.
- *A path toward a muon source for a possible future neutrino factory and, potentially, a muon collider at the Energy Frontier.* This path requires that the new proton source has significant upgrade potential.

These elements are expected to form the basis of the Mission Need statement as is required for Critical Decision 0 (CD-0), and must be incorporated into the design criteria for Project X.

The initial Project X goals and associated design concept [3] were primarily driven by the Project X synergy with the ILC as well as the 2-MW operation of the Main Injector (MI) for the long baseline neutrino program. The details of operation with a slow extracted beam at 8 GeV were not considered in the first proposal. While some enhancements were later introduced in the Project X Initial Configuration Document 1 (ICD-1) it still follows the same path as the first Project X proposal but with an increased beam current. The accelerator complex defined in ICD-1 can drive the long-baseline neutrino program, and provide enhanced capabilities for the mu2e experiment; however, it does not provide a flexible platform to pursue a broader research program in rare muon and kaon processes based on high duty-factor proton beams.

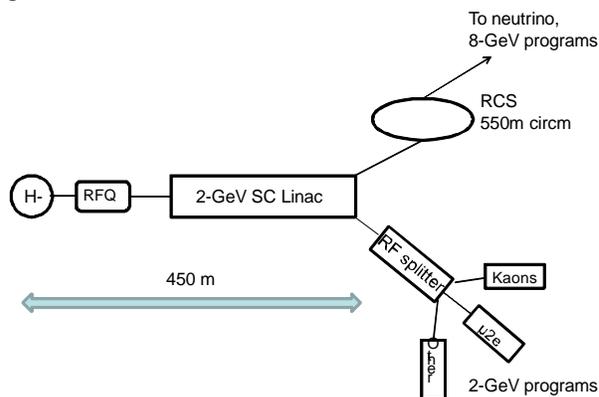

Figure 1: Schematic layout of the ICD-2 proposal.

Motivated by the lack of flexibility in ICD-1, and as part of the standard DOE planning process, the Project-X design team is considering another configuration, named Initial Configuration Document-2 (ICD-2), which can meet the research goals of the expected Mission Need statement.

The ICD-2 is comprised of a 2-GeV superconducting CW (continuous wave) linac, a 2-8 GeV rapid cycling synchrotron (RCS), and the existing (but modified) Recycler and MI to provide 2 MW beam power throughout the energy range 60 – 120 GeV, simultaneous with 2 MW at 2 GeV and 80-190 kW at 8 GeV. A schematic layout is shown in Figure 1. It is anticipated that the final configuration and operating parameters of the complex will be refined through the R&D program in advance of CD-2.

The intent of this configuration is to strengthen the physics program for a wide range of experiments with muons and kaons. It requires a high duty-factor beam with energy above the kaon production threshold (1.6 GeV). In ICD-1 this may be achieved with a slow-extracted beam at 8 GeV. A re-configuration of the low-energy portion of Project X superconducting linac from a pulsed to a CW regime opens new possibilities. Using high efficiency RF separation of the beam allows for a simultaneous operation of several experiments so that each experiment receives the desired beam intensity and structure. In particular, it allows one to run simultaneously two of the highest priority experiments: the mu2e and the rare kaon decays. The requirements of these experiments determine the energy of the CW linac to be at least 2 GeV (kinetic). In addition, depending on the MI operating energy (60 or 120 GeV), 60 to 200 kW of single-turn extracted protons are available from the Recycler at 8 GeV to support experiments, such as g-2, which requires pulsed beam with sufficiently large energy to produce 3.1 GeV muons. In contrast, the high beam power available at 8 GeV in ICD-1 is not readily useable by the rare processes experiments. The slow extraction process has intrinsic limits because of space charge tune-shifts and particle losses; it also typically precludes running simultaneously more than one experiment.

A CW linac provides several important advantages to the rare processes experimental program while preserving the beam characteristics for the long baseline neutrino program. The beam quality and the duty factor of a CW linac are significantly better than that for slow extracted beams. The linac beam intensity does not have the fluctuations inherent in slow extracted beam and has nearly 100% duty factor. The bunch length in a linac (<10 ps rms) is much smaller than can be reasonably achieved in a ring, which allows one to use high accuracy time of flight measurements for particle identification. The beam power in a CW linac is set by high energy physics requirements (ability to use this power by experiments) rather than by technical or accelerator physics limitations. The reduction of particle yield due to decrease of the beam energy from 8 to 2 GeV can be compensated by a higher linac power. But what is more important, the unwanted physics backgrounds tend to decrease (with beam energy) significantly faster than the particle yield for pions and kaons resulting in better overall experimental conditions.

The beam originates from a 1-10 mA DC H$^-$ source (see Figure 1), and is then bunched and accelerated by a CW normal-conducting RFQ to 2.5 MeV. Before entering the main linac the beam is chopped by the bunch-by-bunch chopper following a pre-programmed timeline to create a bunch sequence required by each experiment. From 2.5 MeV to 2 GeV the H$^-$ bunches are accelerated by a CW super-conductive (SCRF) linac. After acceleration the beam is directed to the experiments with subsequent RF separation between them or can be strip injected into the RCS which accelerates the beam to 8 GeV for the accumulation of 6 RCS pulses in the Recycler and subsequent acceleration in the MI.

## 1. RFQ AND BEAM CHOPPER

The RFQ is the only normal conducting accelerating RF structure in the project. Taking into account comparatively small beam current and the necessity of bunch-by-bunch chopping, the RFQ frequency was chosen to be 162.5 MHz. This allows one to achieve the required beam extinction with the state of the art chopping technology. To achieve beam extinction the chopping amplitude has to significantly exceed the beam size. It can be achieved by using several kickers to excite transverse beam motion. It is expected that the total length of the chopping region will be about 5-7 m. It will also include focusing quadrupoles and bunching cavities. The cavities also compress the beam longitudinally to match it to the SC linac operating at 325 MHz.

The chopping pattern and H$^-$ source current are adjusted so that the average beam current in the SC linac would not exceed 1 mA. About 5% of the linac duty cycle (5 ms at 10 Hz) is diverted by a pulsed magnet to a 553-m long RCS with $2.6 \times 10^{13}$ protons per pulse. For the rest of the duty cycle the beam with average power up to 1.9 MW is delivered to the experiments at 2 GeV.

For the optimal linac operation the power of its RF system should be matched to the required beam power. This minimizes the operational cost; and, for constant beam intensity, it results in no energy variations related to the beam intensity. If the average beam intensity stays constant but the peak intensity varies with time so that the beam power (temporarily) exceeds the power of the RF system, the beam energy begins to droop. Fortunately, SC cavities have a comparatively large stored energy which strongly suppresses the energy variations if the beam intensity variations are sufficiently fast. For the accelerating rate of 16 MV/m, suggested for the ILC section of the CW linac, the stored energy in a Tesla-type cavity is ~30 J/cavity. This means that for an average beam current of 1 mA, any intensity redistribution within ~3 μs results in energy gain variation of less than 0.1%. All presently suggested experiments require significantly faster beam intensity variations (or bunching patterns) leading to these variations being "invisible" for the accelerating structures.

## 2. SC LINAC

The CW, 2-GeV linac has an average current (over few microseconds) of 1 mA, with a pulsed current of up to 10 mA. Since the pulsed 8-GeV Project X linac (ICD-1) has a well developed optics operating at this current range, it is possible to use the same structure of the linac and same break points as in the pulsed linac with the necessary modifications to operate in a CW regime. The linac (see Figure 2) consists of a low-energy 325 MHz SCRF section (2.5 - 450 MeV) containing three different families of single-spoke resonators (SSR0, SSR1, SSR2) and one family of a triple-spoke resonator (TSR), and the high energy 1.3-GHz SCRF section (450 MeV – 2 GeV) containing squeezed elliptical $\beta_G$ =0.81 cavities (S-ILC), and ILC-type $\beta_G$=1 cavities.

For this conceptual study we reuse designs of cavities, cryomodules and beam optics (to the extent possible) developed for the pulsed linac in ICD-1. For the ILC-like portion of the CW linac (β=1) we have selected the maximum accelerating cavity gradient of 16 MV/m.

The RMS normalized beam emittance budget is as follows: 0.25 mm-mrad at the ion source, 0.4 mm-mrad at the exit of the CW linac and 0.5 mm-mrad at the injection foil of the RCS. The longitudinal emittance at the linac end is $<5 \cdot 10^{-5}$ eV s.

The CW operation results in a significantly larger heat load on the cryogenic system. The cost optimization (construction + 10 year operation) yields the optimum gradient in the range 16-18 MV/m. We choose 16 MV/m allowing some room to compensate for the loss in acceleration if some cavities are out of operation.

## 3. RAPID CYCLING SYNCHROTRON

To support the 2 MW operation of MI for neutrino program, RCS has to deliver $1.6 \cdot 10^{14}$ particles to Recycler during one MI cycle, which can vary in length from 0.8 sec (60 GeV flattop MI energy) to 1.4 sec (120 GeV flattop MI energy). The RCS has a shorter circumference than Recycler and therefore several RCS cycles are required to fill Recycler. Balancing the impacts of beam space charge, instabilities, magnetic field strength, and repetition rate, the circumference is chosen to be 1/6 of the MI circumference and the repetition rate is chosen to be 10 Hz. During one 0.8 s MI cycle 6 RCS pulses go to MI and the other two are available for an 8 GeV physics program. The main parameters of the RCS are presented in Table 1. The ring acceptance is determined by the Recycler/MI acceptance. To reduce the space charge field of beam accumulated in RCS the small emittance linac beam is painted transversely and longitudinally into larger acceptance of the RCS.

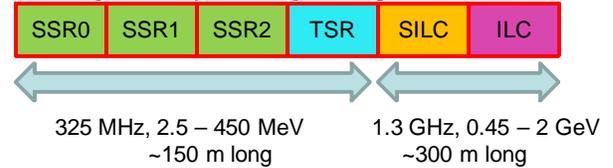

325 MHz, 2.5 – 450 MeV ~150 m long    1.3 GHz, 0.45 – 2 GeV ~300 m long

Figure 2: The schematic of the SCRF linac (2.5 MeV – 2 GeV).

**Table 1: RCS main parameters**

| | |
|---|---|
| Energy, min/max, GeV | 2/8 |
| Repetition rate, Hz | 10 |
| Circumference, m (MI/6) | 553.2 |
| Tunes, $\nu_x/\nu_y$ | 18.42 / 18.44 |
| Transition energy (kinetic), GeV | 13.3 |
| Number of particles | $2.6 \times 10^{13}$ |
| Beam current at injection energy, A | 2.2 |
| $\varepsilon_\perp$ (95% normalized), mm mrad | 22 |
| Space charge tune shift, inj. | 0.07[1] |
| Norm. acceptance at injection, mm mrad | 40 |
| Harmonic number for main RF system, h | 98 |
| RF bucket size at injection, eV s | 0.38 |
| Injection time, ms | 4.3 |
| Required correction of linac energy (kinetic) during injection | 1.2% |
| Total beam power delivered by RCS, kW | 340 |

The ring is designed as a racetrack with FODO lattice of constant periodicity through the entire ring with exception of the injection region. The dispersion is zeroed in the straights by missed dipoles. One long straight section is for the RF cavities and the other is used for injection, extraction, and beam collimation. All F and D quads have the same focusing strength and are connected in series with the dipoles. Eight quadrupoles, four in the injection and four in the extraction regions, have larger aperture and length but the same integral strength. Tune and optics corrections are performed via additional corrector coils wound in each quadrupole. Six out of 132 FODO half-cells have a modified optics obtained by removing 2 quads and by displacing other 4 quads so that the beta-functions at the stripping foil location are increased to about 20 m to reduce the foil heating by the beam during injection (see Figure 3). The betatron phase advance per regular FODO cell is 102°. Strong focusing results in small beam sizes and small dispersion. That, in its turn, results in a small synchrotron beam size and, consequently, a small difference between horizontal and vertical beam envelopes through the entire ring. To ease the power supply voltage requirements the dipoles and quadrupoles of each cell are included into a resonance circuit. Every quadrupole has an associated corrector package which includes a dipole and sextupole coils.

---
[1] This value is computed for the KV-like transverse distribution and the longitudinal bunching factor of 2.2 which are obtained by the beam painting.

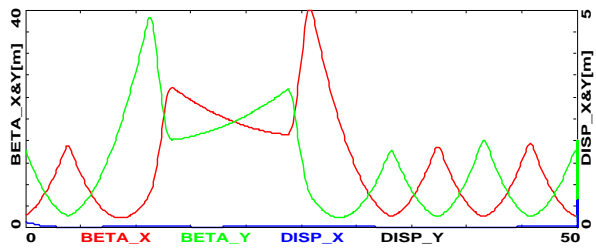

Figure 3: The beta-functions and dispersion in vicinity of injection region

The vacuum chamber for the RCS needs to satisfy a set of opposing requirements. The competing effects are: (1) the shielding and distortion of the dipole bending field by eddy currents excited in the vacuum chamber; (2) the vacuum chamber stability under atmospheric pressure; (3) the vacuum chamber heating by the eddy currents; (4) the transverse impedance due to wall resistivity; (5) and the ring admittance. The compromise resulted in a round stainless steel vacuum chamber with external radius of 22 mm and the wall thickness of 0.7 mm. That results in the vacuum chamber heating by eddy currents of 11 W/m (convective air-cooling with $\Delta T \approx 15$ C°), and the transverse instability with growth rate of the most unstable mode equal to 0.006 turn$^{-1}$. The vacuum chamber acceptance is equal to 40 mm-mrad (normalized, injection energy) with 6 mm allowance for orbit distortions.

The strip injection is produced through the 420 μg/cm$^2$ foil as shown in Figure 4. The beam painting is achieved by a closed orbit displacement in the horizontal and vertical planes resulting in ~40 foil passages per particle for 2200 turn injection. The increase of beta-functions on the foil relative to their values of FODO lattice decreased the density of particle flux from 7 to 2 mm$^{-2}$ per particle with corresponding reduction of foil heating.. The foil is mainly cooled by the black body radiation. To make the radiative cooling more effective the foil is tilted by 45 deg. relative to the beam direction. It results in a desired effective foil thickness of 600 μg/cm$^2$ and yields the stripping efficiency better than 99%. A foil temperature estimate, which takes into account only radiative cooling and heating reduction due to δ-electrons, results in the peak value of 1500 K°.

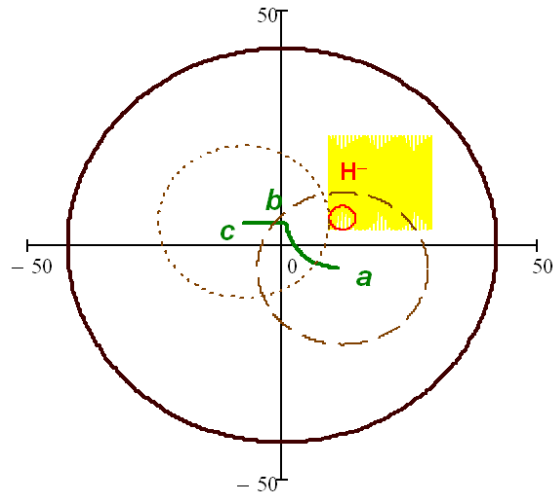

Figure 4: The beam and vacuum chamber cross-sections (in mm) at the foil location. The green line shows displacement of the closed orbit in the course of painting. It starts at point *b*, goes to point *a*, then goes back to *b*, and finally it is moved to point *c* to prevent further beam interaction with the foil. The yellow square shows position of the stripping foil. The red ellipse shows the boundary of the injected H$^-$ beam. The brown dashed and dotted lines present the boundaries of the stored proton beam when the closed orbit is located at points *a* and *c* for the machine acceptance of 40 mm-mrad (normalized). The internal radius of the vacuum chamber (outer circle) is 42 mm.

## 4. UPGRADE SCENARIOS

There are two main avenues for the ICD-2 upgrade for the muon collider or neutrino factory. The first one is based on an upgrade of RCS; while the second one is based on a construction of extension to the CW linac. Both proposals require a new compressor ring to meet the NF/MC bunch requirements. In this document we assume that this ring energy is 8 GeV. Taking into account that both the neutrino factory and the muon collider require pulsed beams the linac extension should be based on a pulsed SC linac. There is also a possibility of combined approach where a linac extension and a new higher energy rapid cycling synchrotron are constructed. While we will not consider this case an application of the discussed below limitations is straight forward.

In further consideration we also assume that an upgrade is build shortly after commissioning of ICD-2 accelerator complex and therefore should not significantly affect the discussed above physics program.

First, let us consider the case when the ICD-2 complex is used for muon production without any upgrade. To compress the bunch to the rms bunch length of 60 cm the longitudinal 95% beam emittance in the compressor ring should not exceed 3.5 eV s; *i.e.* not more than 10 RCS bunches can be coalesced. However, taking into account that the transverse emittance of the beam in the compressor ring can be much larger than the RCS beam emittance, one can do a multi-turn injection in the horizontal plane. For the compressor ring 3 times shorter than RCS one can do 3 turn injection and, consequently, can coalesce 30 RSC bunches. That implies that three trains of 10 bunches each has to be injected to the RCS. After acceleration they are extracted in one turn from RCS and are injected to the compressor ring with the three-turn injection resulting in 10 bunches in the compressor ring. After adiabatic coalescing and bunch rotation one obtains a single bunch with ~60 cm rms length. If RCS delivers all its beam power to the compressor ring, the total beam power is 340 kW. In this case the only limitation to be addressed in the entire ICD-2 complex is a somewhat higher beam loading due to a factor of ~3 larger single bunch intensity. An increased value of the space charge tune shift (~0.2) looks manageable (still less than present FNAL Booster).

Further increase of the beam power requires significant and expensive hardware changes.

The most straightforward RCS upgrade would be an increase of RCS repetition rate to 20 Hz. That results in the following changes. First, it requires a change in resonance circuits of bending magnets and quads. Second, the 4 times increase of vacuum chamber heating by eddy current requires its forced air-cooling. Third, the 2 times faster accelerating rate requires twice more accelerating cavities and an upgrade of low level RF. In this case RF cavities will fill all available space in the ring. Fourth, it requires doubling of linac current. Its implications are discussed below. Although twice as many particles are strip-injected, the foil peak temperature is slightly decreased because the injection process is twice as fast. All of that would result in the maximum beam power of ~700 kW.

For the considered above parameters the space charge tune shift is already ~0.2 and therefore further beam current increase would require an increase of beam size and, consequently, vacuum chamber aperture. The beam current increase of about 2 times can be potentially achieved resulting in ~1.4 MW beam power at 20 Hz. Such an upgrade will require: (1) an increase of horizontal aperture of the vacuum chamber requiring improvements in its forced air cooling, (2) doubling the RF power and (3) an increase of linac beam current to 4 mA. The aperture increase will also reduce the foil heating which otherwise would be a potential showstopper. Note that a replacement of a foil striping by the laser striping is problematic for RCS due to ~1% energy change during injection. It requires corresponding changes of the laser wavelength which is not a trivial problem for the required effective laser power (≥100 kW).

An increase of the linac beam current requires a proportional increase of linac RF power. A direct increase of CW RF power is expensive and does not look prudent. More attractive approach is to combine power of CW and pulsed RF sources. Conceptually, it is straightforward [4] but requires an extensive R&D to demonstrate its operation on the required power level.

A replacement of acceleration in RCS by acceleration in a pulsed linac can significantly reduce risks associated with beam losses and acceleration in RCS; but for a bunch frequency of 20 Hz and 8 GeV energy it still does not allow to exceed the RCS power significantly. It is related to the limitations on the longitudinal and transverse beam emittances [5]. However, a pulsed linac can operate at much higher repetition frequency than RCS resulting in a proportional beam power increase.

For the case of limited bunch repetition rate (≤20 Hz) a further beam power increase will require higher beam energy. Both the synchrotron and linac choices can be used while the synchrotron option looks like a less expensive option.

## ACKNOWLEDGMENTS

We would like to thank the entire Project X design team for contributing to both ICD-1 and ICD-2. Specifically, we are thankful to Steven Holmes and Paul Derwent for numerous discussions on Project X beam requirements and operation scenarios.